\documentclass[twocolumn,showpacs,showkeys,preprintnumbers,amsmath,amssymb]{revtex4-1}
\usepackage{bbm}
\usepackage{amsfonts}

\usepackage{epsfig}
\usepackage{graphicx}
\usepackage{amssymb}   
\usepackage{dcolumn}
\usepackage{bm}
\usepackage[colorlinks,citecolor=blue, linkcolor=blue,hyperindex]{hyperref}
\begin{document}
\vspace{2.5cm}
\title{Retrieving the lost fermionic entanglement by partial measurement in noninertial frames}
\author{Xing Xiao$^{1}$}
\author{Ying-Mao Xie$^{1}$}
\altaffiliation{xieyingmao@126.com}
\author{Yao Yao$^{2}$}
\author{Yan-Ling Li$^{3}$}
\altaffiliation{liyanling0423@gmail.com}
\author{Jieci Wang$^{4}$}
\altaffiliation{jcwang@hunnu.edu.cn}
\affiliation{$^{1}$College of Physics and Electronic Information,
Gannan Normal University, Ganzhou Jiangxi 341000, China\\
$^{2}$Microsystems and Terahertz Research Center, China Academy of Engineering Physics, Chengdu, Sichuan 610200, China\\
$^{3}$School of Information Engineering, Jiangxi University of Science and Technology, Ganzhou, Jiangxi 341000, China\\
$^{4}$Institute of Physics and Department of Physics, Hunan Normal University, Changsha, Hunan 410081, P. R. China}

\begin{abstract}
The initial entanglement shared between inertial and accelerated observers degrades due to the influence of the Unruh effect. Here, we show that the Unruh effect can be completely eliminated by the technique of partial measurement. The lost entanglement could be entirely retrieved or even amplified, which is dependent on whether the optimal strength of reversed measurement is \emph{state-independent} or \emph{state-dependent}. 
Our work provides a novel and unexpected method to recover the lost entanglement under Unruh decoherence and exhibits the ability of partial measurement as an important technique in relativistic quantum information.
\end{abstract}
\maketitle

\section{Introduction}

Relativistic quantum information \cite{alsing2012}, a combination of relativity theory and quantum information, is a rapidly developing new field of physics. It aims to blend together the concepts from relativity and quantum information, and understand them from each other's perspective. The latest research shows that a deeper comprehension of the quantum entanglement (or quantum correlations) in a relativistic frame is not only of interest to quantum information processing \cite{peres2004,bradler2007}, but also plays a crucial role in understanding black hole thermodynamics \cite{bombelli1986,callan1994} and information paradox of black holes \cite{hawking1974,hawking1976,zhang2009,tera2000,zhang2011,zhang2013}. Another exciting development in this field is that several quantum information experiments, such as the quantum communications between the Earth and satellites, are already approaching relativistic regimes \cite{wilson2011,rideout2012,friis2013,laht2013,vallone2016}.

In relativistic quantum information, one of the most distinguished topic is how the entanglement and entanglement-involved quantum information protocols are affected by Unruh-Hawking effects \cite{hawking1974,unruh1976}. There are many literatures showing that entanglement is
dependent on the observer's motion \cite{schuller2005,emm2010,alsing2006,leon2009,emm2009,dcm2009,bruschi2010,wang2010,xiao2011,wang2011,dai2015,richter2015}. When one of the partners of an entangled system undergoes uniform acceleration, the initial entanglement between the partners is degraded. The degradation of entanglement inevitably reduces the confidence of entanglement-based quantum information tasks (e.g., quantum teleportation \cite{alsing2003,alsing2004,pan2008}).
This phenomenon is directly related to the Unruh effect and usually known as Unruh decoherence. Particularly, the results show significant differences between bosonic and fermionic fields: the bosonic field entanglement vanishes asymptotically at the infinite acceleration limit 
while the fermionic field entanglement is found never to be completely destroyed. This notable difference is usually attributed to the fermionic/bosonic statistics \cite{alsing2006}. Although it was widely believed that the Unruh effect can only cause the degradation of entanglement shared between an inertial and an accelerated observer, the realistic results are more subtle. Montero and Mart\'in-Mart\'inez found that beyond the single-mode approximation (SMA) and for some particular initial states, net entanglement between inertial and accelerated observers could be created by the Unruh effect \cite{montero2011}. However, the created entanglement is rather limit and makes little contribution to relativistic quantum information processing. Note that the aforementioned discussions are restricted to consider the entanglement degradation in noninertial frames, while limited attention is paid to the protection or retrieval of the lost entanglement against Unruh decoherence.

In this paper, we propose a scheme to retrieve the lost fermionic entanglement by partial measurement in noninertial frames. Partial measurement (or weak measurement in some references \cite{ueda1992,royer1994,koro2010}) is associated with the positive-operator valued measure (POVM). By ``partial'', we mean that the wave-function doesn't completely collapse under this type of measurement. Hence, when needed, the initial state could be restored with some operations. In non-relativistic quantum information, partial measurement has been extensively demonstrated to protect the quantum entanglement from amplitude damping decoherence (energy relaxation from the excited state of a two-level system to its ground state) both theoretically \cite{sun2010,man2012,xiao2013} and experimentally \cite{koro2006,katz2008,lee2011,kim2012}. The key idea of these proposals for entanglement protection is based on the fact that partial measurement is not completely destructive and can be reversed by a series of operations named as partial measurement reversal with a certain probability. Here, we borrow this technique from quantum information and show that the lost fermionic entanglement could be retrieved from Unruh decoherence by utilizing partial measurement. Our proposal is universal for arbitrary initial states and also works for other types of quantum correlations, e.g., quantum discord \cite{vedral2001,zurek2001}.


The remainder of this paper is organized as follows. First, in section \ref{sec2.1}, we review the essential results of Unruh effect for Dirac particles as experienced by a uniformly accelerating observer named Rob. Despite the details could be found in ref. \cite{alsing2006}, we point out that the Unruh decoherence for Rob is is equivalent to the \emph{anti-amplitue damping decoherence}. Consequently, it could be reformulated mathematically by Kraus operators. Then in section \ref{sec2.2}, we construct the proper form of partial measurement and partial measurement reversal to battle against the Unruh decoherence. Section \ref{sec3} discusses the recovery of entanglement with the help of partial measurement. Particularly, we show two methods (i.e., \emph{state-independent} and \emph{state-dependent}) to obtain the optimal strength of partial measurement reversal, which are manifested in section \ref{sec3.1} and section \ref{sec3.2}, respectively. Furthermore, we demonstrate that quantum discord also can be retrieved by the same procedure, as shown in section \ref{sec4}. Finally, we summarize our results in section \ref{sec5}.

\begin{figure}[hbtp] \centering
\includegraphics[width=0.5\textwidth]{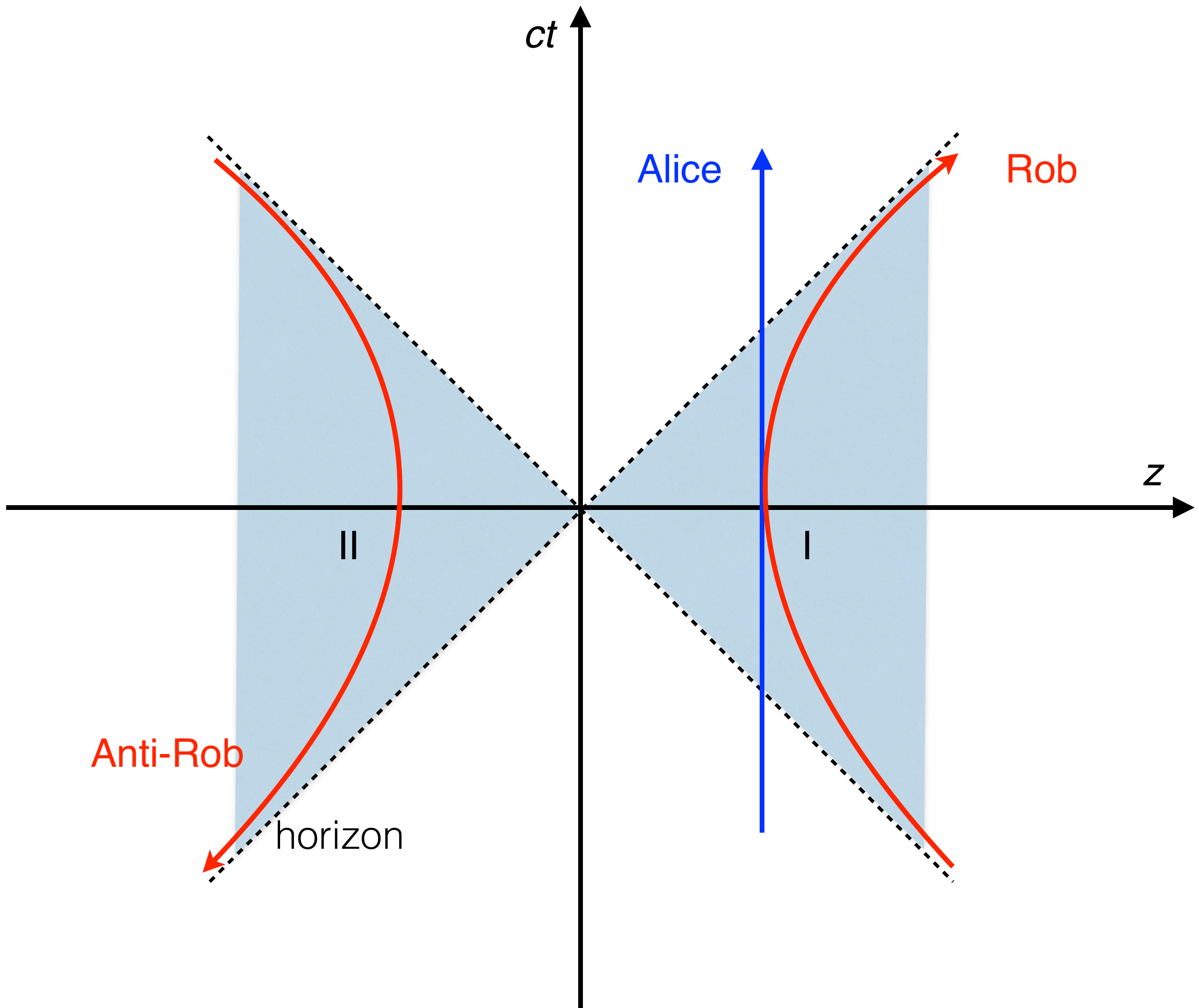}
\caption{(color online) Diagram showing Rindler spacetime (with the two perpendicular space dimensions suppressed) embedded into Minkowski spacetime. The straight line shows the world line of an inertial observer Alice (A). An uniformly accelerated observer Rob (R) with acceleration $a$ travels on a hyperbolic world line constrained to region I, while a fictitious observer antiRob (R) travels on the corresponding hyperbola in region II. Alice and Rob share an entangled Bell state at the beginning. Minkowski spacetime diagram showing the world lines of an inertial observer Alice, and one uniformly accelerated observer that can be moving in either the Rindler wedge I or II which are causally disconnected from each other.}
\label{Fig1}
\end{figure}

\section{Preliminaries}
\label{sec2}
\subsection{Quantization of the Dirac field}
\label{sec2.1}
Let us consider a free Minkowski Dirac field $\psi$ in 1+1 dimensions
\begin{equation}
\label{eq2.1}
i\gamma^{\mu}\partial_{\mu}\psi-m\psi=0,
\end{equation}
where $m$ is the particle mass, $\gamma^{\mu}$ are the Dirac gamma matrices, and $\psi$ is a spinor wave function. The field can be expressed from the perspective of inertial and uniformly accelerated observers, which has been investigated in detail in \cite{alsing2006}. Here we review the main results that are related to our following discussions.

The system of our scheme includes an inertial observer, Alice (A) and a uniformly accelerated observer, Rob (R). For the inertial observer, Minkowski coordinates $(t,z)$ are the most appropriate choice of coordinates to describe the field. Then the field can be quantized by expanding it in terms of a complete orthonormal set of positive- and negative- frequency Minkowski modes $\psi_{\mathbf{k}}^{+}$ and $\psi_{\mathbf{k}}^{-}$
\begin{equation}
\label{eq2.2}
\psi=\int d\mathbf{k}\left(a_{\mathbf{k}}\psi_{\mathbf{k}}^{+}+b_{\mathbf{k}}^{\dagger}\psi_{\mathbf{k}}^{-}\right)
\end{equation}
where the wave vector $\mathbf{k}$ denotes the modes of massive Dirac fields. $a_{\mathbf{k}}^{\dagger}$, $b_{\mathbf{k}}^{\dagger}$ and $a_{\mathbf{k}}$, $b_{\mathbf{k}}$ are the creation and annihilation operators for the positive- and negative- frequency modes of momentum $\mathbf{k}$ that satisfy the anticommutation relations $\{a_{\mathbf{k}},a_{\mathbf{k}'}^{\dagger}\}=\{b_{\mathbf{k}},b_{\mathbf{k}'}^{\dagger}\}=\delta_{\mathbf{k}\mathbf{k}'}$
with all other anticommutators vanishing. 

However, for the uniformly accelerated observer, Rindler
coordinates $(\tau,\xi)$ should be employed for describing the viewpoint of a non-inertial observer. As shown in Fig. \ref{Fig1}, due to the eternal acceleration, Rob travels on a hyperbola constrained in the Rindler region I which is causally disconnected from region II. The coordinate transformation between Minkowski coordinates $(t,z)$ and Rindler coordinates $(\tau,\xi)$ is given by
$at=e^{a\xi}\sinh(a\tau)$, $az=e^{a\xi}\cosh(a\tau)$. In analogy to Eq. (\ref{eq2.2}), the Dirac field in terms of the complete set of positive- and negative- frequency Rindler modes (solutions of the Dirac equation in Rindler coordinates) is given by
\begin{equation}
\label{eq2.3}
\psi=\int d\mathbf{k}\left(c_{\mathbf{k}}^{\rm I}\psi_{\mathbf{k}}^{\rm I+}+d_{\mathbf{k}}^{\rm I\dagger}\psi_{\mathbf{k}}^{\rm I-}+c_{\mathbf{k}}^{\rm II}\psi_{\mathbf{k}}^{\rm II+}+d_{\mathbf{k}}^{\rm II\dagger}\psi_{\mathbf{k}}^{\rm II-}\right),
\end{equation}
where $(c_{\mathbf{k}}^{n},c_{\mathbf{k}}^{n\dagger})$ and $(d_{\mathbf{k}}^{n},d_{\mathbf{k}}^{n\dagger})$ denote the annihilation and creation operators for the Rindler particle and antiparticle, respectively, in the region $n$ with $n=\rm I$, $\rm II$. They obey the usual Dirac anticommutation relations $\{c_{\mathbf{k}}^{n},c_{\mathbf{k}'}^{n'\dagger}\}=\{d_{\mathbf{k}}^{n},d_{\mathbf{k}'}^{n'\dagger}\}=\delta_{nn'}\delta_{\mathbf{k}\mathbf{k}'}$.

The Minkowski and Rindler creation and annihilation operators is connected by the Bogoliubov transformation
\begin{equation}
\label{eq2.4}
a_{\mathbf{k}}=\cos rc_{\mathbf{k}}^{\rm I}-\sin rd_{-\mathbf{k}}^{\rm II\dagger},\ \  b_{\mathbf{-k}}^{\dagger}=\sin rc_{\mathbf{k}}^{\rm I}+\cos rd_{-\mathbf{k}}^{\rm II\dagger},
\end{equation}
where $r=\arccos\sqrt{1+e^{-2\pi\omega/a}}$ with $a$ the Rob's acceleration. Note that $r\in[0,\pi/4]$ as $a$ ranges from 0 to infinity. It is easy to see from Eq. (\ref{eq2.4}) that the Minkowski annihilation operator $a_{\mathbf{k}}$ mixes the Rindler annihilation operator $c_{\mathbf{k}}^{\rm I}$ in region \rm{I} of momentum $\mathbf{k}$ and the Rindler creation operator $d_{-\mathbf{k}}^{\rm II\dagger}$ in region \rm{II} of momentum $-\mathbf{k}$. This mixture indicates that the Unruh vacuum could be expressed in terms of the Rindler region \rm{I} and region \rm{II} states
\begin{equation}
\label{eq2.5}
|0_{\mathbf{k}}\rangle_{U}=\cos r|0_{\mathbf{k}}\rangle_{\rm I}|0_{-\mathbf{k}}\rangle_{\rm II}+\sin r|1_{\mathbf{k}}\rangle_{\rm I}|1_{-\mathbf{k}}\rangle_{\rm II}.
\end{equation} 
Similarly, the one-particle state is given by
\begin{equation}
\label{eq2.6}
|1_{\mathbf{k}}\rangle_{U}=|1_{\mathbf{k}}\rangle_{\rm I}|0_{-\mathbf{k}}\rangle_{\rm II}.
\end{equation}
Note that the observers in Rindler region \rm{I} and region \rm{II} are causally disconnected from each other.
For Rob in region \rm{I}, the mode \rm{II} should be traced out since it is not observable. Therefore, the total effect of Unruh decoherence for a uniformly accelerated observer could be reformulated mathematically as a completely positive trace-preserving (CPTP) linear map.
The corresponding Kraus operators are given as
\begin{eqnarray}
\label{eq2.7}
E_{1}=\left(\begin{array}{cc}\cos r & 0 \\0 & 1\end{array}\right), 
E_{2}=\left(\begin{array}{cc}0 & 0 \\\sin r & 0\end{array}\right).
\end{eqnarray}
It is straightforward to find that the form of Eq. (\ref{eq2.7}) extremely resembles the amplitude damping decoherence which has been well understood in quantum optics and quantum information theory \cite{nielsen2000}. The only difference is the inverse arrangement of the elements of matrix. Thus, we call Eq. (\ref{eq2.7}) as \emph{anti-amplitude damping decoherence}. 

With the above Eqs. (\ref{eq2.5}) and (\ref{eq2.6}) in mind, one can analyze the degradation of entanglement from the perspective of observers in uniform acceleration (cf. refs. \cite{alsing2006,wang2010,xiao2011,wang2011}). However, 
in this paper, we show that the lost fermionic entanglement can be recovered in a probabilistic way via partial measurement and its reversal. The main idea of our scheme is illustrated in Fig. \ref{Fig1}, where we have considered the following scenario: (i) Alice (blue arrow) stays stationary, while Rob (red hyperbola) undergoes constant acceleration. Alice and Rob share an entangled state at the beginning. (ii) A partial measurement is performed on Rob before its acceleration (i.e, at time $\tau=0$). (iii) After Rob's acceleration (assuming Rob has instantaneously accelerated to the value $a$), the operation of partial measurement reversal is implemented by Rob in the region \rm I.

\subsection{Partial measurement and partial measurement reversal}
\label{sec2.2}
Before discussions, let us first introduce the partial measurement and partial measurement reversal briefly. In contrast to the standard von Neumann projective measurement, which projects the initial state to one of the eigenstates of the measurement operator. Partial measurement is a generalization of the standard von Neumann projective measurement, usually known as POVM. The most significant advantage of partial measurement is that it doesn't completely collapse the initial state, thereby keeping the measured state reversible. The partial measurement could be formally written as
\begin{equation}
\label{eq2.8}
\mathcal{M}_{0}=\sqrt{1-p}|0\rangle\langle0|+|1\rangle\langle1|,
\end{equation}
where $\mathcal{M}_{0}^{\dagger}\mathcal{M}_{0}+\mathcal{M}_{1}^{\dagger}\mathcal{M}_{1}=I$ with $\mathcal{M}_{1}=\sqrt{p}|0\rangle\langle0|$. The parameter $p$ $(0\leqslant p\leqslant1)$ is usually named as the strength of partial measurement. Note that $\mathcal{M}_{0}$ and $\mathcal{M}_{1}$ are not necessarily projectors and also nonorthogonal to each other.
One may note that the partial measurement described by Eq. (\ref{eq2.8}) slightly differs from the well-informed expression of partial measurement which is discussed in Refs. \cite{koro2010,sun2010,man2012,xiao2013,koro2006,katz2008,lee2011,kim2012}. This divergence stems from the different decoherence mechanism between amplitude damping decoherence and Unruh decoherence (\emph{anti-damping amplitude decoherence}).

As one may expect, the more information is obtained from a quantum system, the more its state is disturbed by measurement. Intriguingly, the partial measurement of Eq. (\ref{eq2.8}) only induces a partial collapse of quantum state, hence the initial state can be retrieved with a nonzero success probability by reversing operations on the post-measurement state \cite{cheong2012}.
The procedure of reversal can be described by a non-unitary operator
\begin{eqnarray}
\label{eq2.9}
\mathcal{M}_{0}^{-1}&=&\frac{1}{\sqrt{1-q}}\left(\begin{array}{cc}0 & 1 \\1 & 0\end{array}\right)\left(\begin{array}{cc}\sqrt{1-q} & 0 \\0 & 1\end{array}\right)\left(\begin{array}{cc}0 & 1 \\1 & 0\end{array}\right)\\
&=&\frac{1}{\sqrt{1-q}}X\mathcal{M}_{0}(q)X,\nonumber
\end{eqnarray}
where $X=|0\rangle\langle1|+|1\rangle\langle0|$ is the bit-flip operation (i.e., Pauli X operation) and $\mathcal{M}_{0}(q)$ denotes the same form of Eq. (\ref{eq2.8}) by replacing $p$ with $q$. Note that the Eq. (\ref{eq2.9}) does not represent a measurement operator, but a reversing process. Although the non-unitary operator $\mathcal{M}_{0}^{-1}$ cannot be realized as an evolution with a suitable Hamiltonian, it can be achieved with certain probability by using another partial measurement. The second line of Eq. (\ref{eq2.9}) indicates that the reversing process could be constructed by three steps: a bit-flip operation, a second partial measurement and a second bit-flip operation.
The factor $1/\sqrt{1-q}$ is related to the non-unitary nature of the partial measurement, i.e., probabilistic.
Therefore, the partial measurement reversal could exactly undo the partial measurement $\mathcal{M}_{0}$ in a probabilistic way by properly choosing the parameter $q$.

\section{Retrieving the lost fermionic entanglement}
\label{sec3}
Let Alice and Rob initially share the entangled state
\begin{equation}
\label{eq3.1}
|\Psi(0)\rangle=\alpha|0\rangle_{A}|0\rangle_{R}+\beta|1\rangle_{A}|1\rangle_{R},
\end{equation}
where $\alpha^2+\beta^2=1$, $\alpha$ and $\beta$ are real numbers. After the coincidence of Alice and Rob, Rob performs a partial measurement of (\ref{eq2.8}) on his own particle. If the partial measurement is successfully carried out (with the probability $1-p$), the state of (\ref{eq3.1}) reduces to
\begin{equation}
\label{eq3.2}
|\Psi(1)\rangle=\frac{1}{\sqrt{N_{1}}}\left(\alpha\sqrt{\overline{p}}|0\rangle_{A}|0\rangle_{R}+\beta|1\rangle_{A}|1\rangle_{R}\right),
\end{equation}
where $N_{1}=\alpha^2\overline{p}+\beta^2$ is the normalization factor and $\overline{p}=1-p$. Note that
the amplitude of the state $|1\rangle_{A}|1\rangle_{R}$ is amplified due to the shrink of the normalization factor. As $p$ approaches unity, the state $|\Psi(1)\rangle$ is almost completely projected into the state $|1\rangle_{A}|1\rangle_{R}$, which is free from the Unruh effect.

Given that Rob undergoes uniform acceleration, the states $|0\rangle_{R}$ and $|1\rangle_{R}$ of Rob should be expanded as Eqs. (\ref{eq2.5}) and (\ref{eq2.6}). Thus the state $|\Psi(1)\rangle$ changes to
\begin{eqnarray}
\label{eq3.3}
|\Psi(2)\rangle=&&\frac{1}{\sqrt{N_{1}}}\Big[\alpha\sqrt{\overline{p}}\big(\cos r|0\rangle_{A}|0\rangle_{\rm I}|0\rangle_{\rm II}\\
&&+\sin r|0\rangle_{A}|1\rangle_{\rm I}|1\rangle_{\rm II}\big)+\beta|1\rangle_{A}|1\rangle_{\rm I}|0\rangle_{\rm II}\Big],\nonumber
\end{eqnarray}
where the acceleration parameter $r$ ranges from 0 to $\pi/4$ since $0<a<\infty$. As we illustrated in Fig.
\ref{Fig1}, in order to remove the Unruh effect and recover the initial entanglement, a partial measurement reversal is performed by Rob in the region \rm I. After the successful performance of partial measurement reversal, we obtain
\begin{eqnarray}
\label{eq3.4}
|\Psi(3)\rangle=&&\frac{1}{\sqrt{N_{2}}}\Big[\alpha\sqrt{\overline{p}}\big(\cos r|0\rangle_{A}|0\rangle_{\rm I}|0\rangle_{\rm II}\\
&&+\sin r\sqrt{\overline{q}}|0\rangle_{A}|1\rangle_{\rm I}|1\rangle_{\rm II}\big)+\beta\sqrt{\overline{q}}|1\rangle_{A}|1\rangle_{\rm I}|0\rangle_{\rm II}\Big],\nonumber
\end{eqnarray}
with the normalization factor $N_{2}=\alpha^2\overline{p}\cos^{2}r+\alpha^2\overline{p}\overline{q}\sin^{2}r+\beta^2\overline{q}$ and $\overline{q}=1-q$. $q$ is the strength of the second partial measurement.
Since Rob is causally disconnected from region II, we must trace over the mode \rm II, which results in a mixed state between Alice and Rob,
\begin{eqnarray}
\label{eq3.5}
\rho_{A,\rm I}=\frac{1}{N_{2}}\left[\begin{array}{cccc}\alpha^2\overline{p}\cos^{2}r & 0 & 0 & \alpha\beta\sqrt{\overline{p}\overline{q}}\cos r \\0 & \alpha^2\overline{p}\overline{q}\sin^{2}r & 0 & 0 \\0 & 0 & 0 & 0 \\\alpha\beta\sqrt{\overline{p}\overline{q}}\cos r & 0 & 0 & \beta^2\overline{q}\end{array}\right]\nonumber\\
\end{eqnarray}

\subsection{State-independent optimal partial measurement reversal}
\label{sec3.1}
Let us now demonstrate how the partial measurement retrieves the lost fermionic entanglement. In order to quantify the shared entanglement between Alice and Rob, we
use the concurrence \cite{wootters1998} to measure the entanglement. It is
defined as
\begin{equation}
\label{eq3.6}
C=\max\{0,\sqrt{\lambda_{1}}-\sqrt{\lambda_{2}}-\sqrt{\lambda_{3}}-\sqrt{\lambda_{4}}\},
\end {equation}
where $\lambda_{i},(i=1,2,3,4)$ are the eigenvalues, in decreasing
order, of matrix
$\tilde{\rho}=\rho(\sigma_{y}\otimes\sigma_{y})\rho^*(\sigma_{y}\otimes\sigma_{y})$;
$\sigma_{y}$ is the Pauli spin matrix and the asterisk denotes
complex conjugation. The concurrence varies from $C=0$ for a
separable state to $C=1$ for a maximally entangled state. In
particular, the concurrence, for the density matrix of Eq. (\ref{eq3.5}), is
given by
\begin{equation}
\label{eq3.7}
C_{\rm PM}=2\alpha\beta\sqrt{\overline{p}\overline{q}}\cos r/N_{2}.
\end{equation}
It is straightforward to note that if no partial measurement and its reversal is involved (i.e., $p=q=1$ and then $N_{2}=1$), our result immediately reverts to 
\begin{equation}
\label{eq3.8}
C_{\rm UD}=2\alpha\beta\cos r,
\end{equation}
which is the concurrence under pure Unruh decoherence, as shown in Ref.\cite{alsing2006}. 

To achieve the maximum amount of entanglement between Alice and Rob,  we have to choose the optimal reversing measurement strength $q$. An intuitive idea is to ensure that the final state is, wherever possible,
close to the initial state.The quantum jump approach offers a solution (see details in the appendix). The state-independent optimal reversing measurement strength is given as
\begin{equation}
\label{eq3.9}
q_{\rm SI}=1-(1-p)\cos^2r.
\end{equation}
where the subscript $\rm SI$ indicates the state-independent case. It is remarkable to notice that Eq. (\ref{eq3.9}) is concise and state-independent (i.e., without involving the initial parameters $\alpha$ and $\beta$), which drastically releases the constraint conditions on experimental tests.

Consequently, under the state-independent optimal partial measurement reversal, the maximally retrievable concurrence $C_{\rm SI}^{\rm opt}$ turns out to be 
\begin{equation}
\label{eq3.10}
C_{\rm SI}^{\rm opt}=\frac{2\alpha\beta}{1+\alpha^2\overline{p}\sin^2r}.
\end{equation}
From Eq. (\ref{eq3.10}), we can draw an important conclusion that the strength of partial measurement $p$ plays a crucial role in the maximally retrievable concurrence $C_{\rm SI}^{\rm opt}$. 
\begin{figure*}[hbtp] \centering
\includegraphics[width=0.9\textwidth]{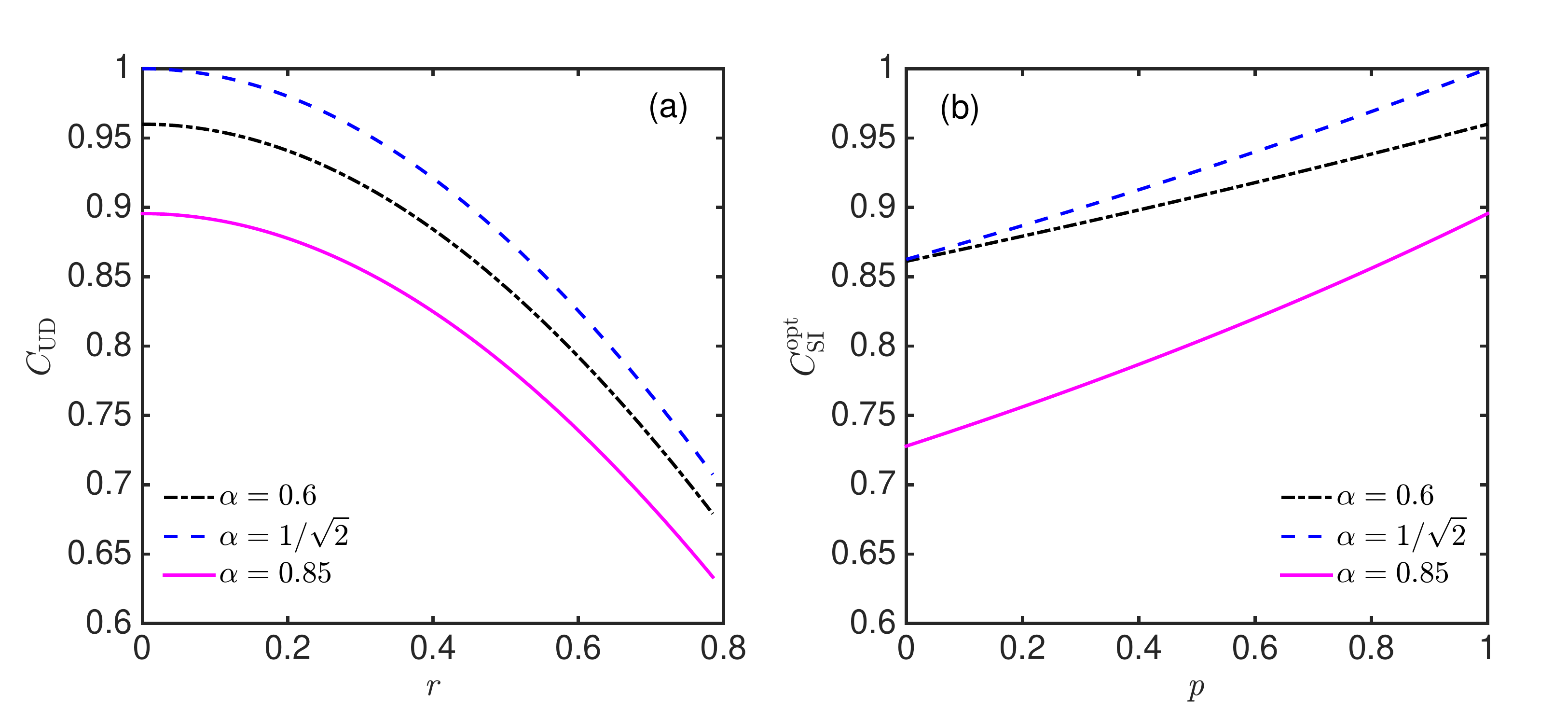}
\caption{(color online) (a). The Unruh decohered concurrence $C_{\rm UD}$ as a function of the acceleration parameter $r$. (b) The maximally retrievable concurrence $C_{\rm SI}^{\rm opt}$ as a function of the partial measurement strength $p$, where we have chosen the acceleration parameter $r=0.6$ and the state-independent optimal reversing measurement strength $q_{\rm SI}=1-(1-p)\cos^2r$.}
\label{Fig2}
\end{figure*}
 
A careful comparison with the Unruh decohered concurrence $C_{\rm UD}$ and the reversed concurrence $C_{\rm SI}^{\rm opt}$ is presented in Fig. \ref{Fig2}.
Unlike the results in Fig. \ref{Fig2}(a) where the monotonic degradation of entanglement occurs due to the Unruh decoherence, Fig. \ref{Fig2}(b) shows that partial measurement and partial measurement reversal can be indeed used for recovering the lost fermionic entanglement from decoherence. 
With given values of initial parameter $\alpha$ and accelerated parameter $r$, the larger $p$ is, the more entanglement is retrieved. Particularly, in the limit $p\rightarrow1$, the initial entanglement shared between Alice and Rob could be almost completely retrieved. Moreover, according to Eq. (\ref{eq3.10}), we notice that this conclusion is always valid regardless of the initial parameter $\alpha$ and the acceleration parameter $r$. 
 

Since the partial measurement is not a unitary operation, the recovery of the lost entanglement is not deterministic but probabilistic. In the case of state-independent optimal partial measurement reversal, the success probability is
\begin{equation}
\label{eq3.11}
\mathcal{P}_{\rm SI}^{\rm opt}=\overline{p}\cos^2r(1+\alpha^2\overline{p}\sin^2r).
\end{equation}

In fact, a partial measurement is equivalent to a probabilistic state rotation with the rotation angle corresponding to the strength of the measurement. Therefore, the first partial measurement is intentionally performed to project Rob's state towards the `lethargic' state $|1\rangle_{R}$ which is immune to the Unruh decoherence. The closer the initial state is projected towards the `lethargic' state, the larger the amount of reversed entanglement is finally obtained. Meanwhile, the success probability decreases. Thus, the trade-off between the large retrieved concurrence and low success probability should be carefully optimized in realistic scenario.

\begin{figure*}[hbtp] \centering
\includegraphics[width=0.9\textwidth]{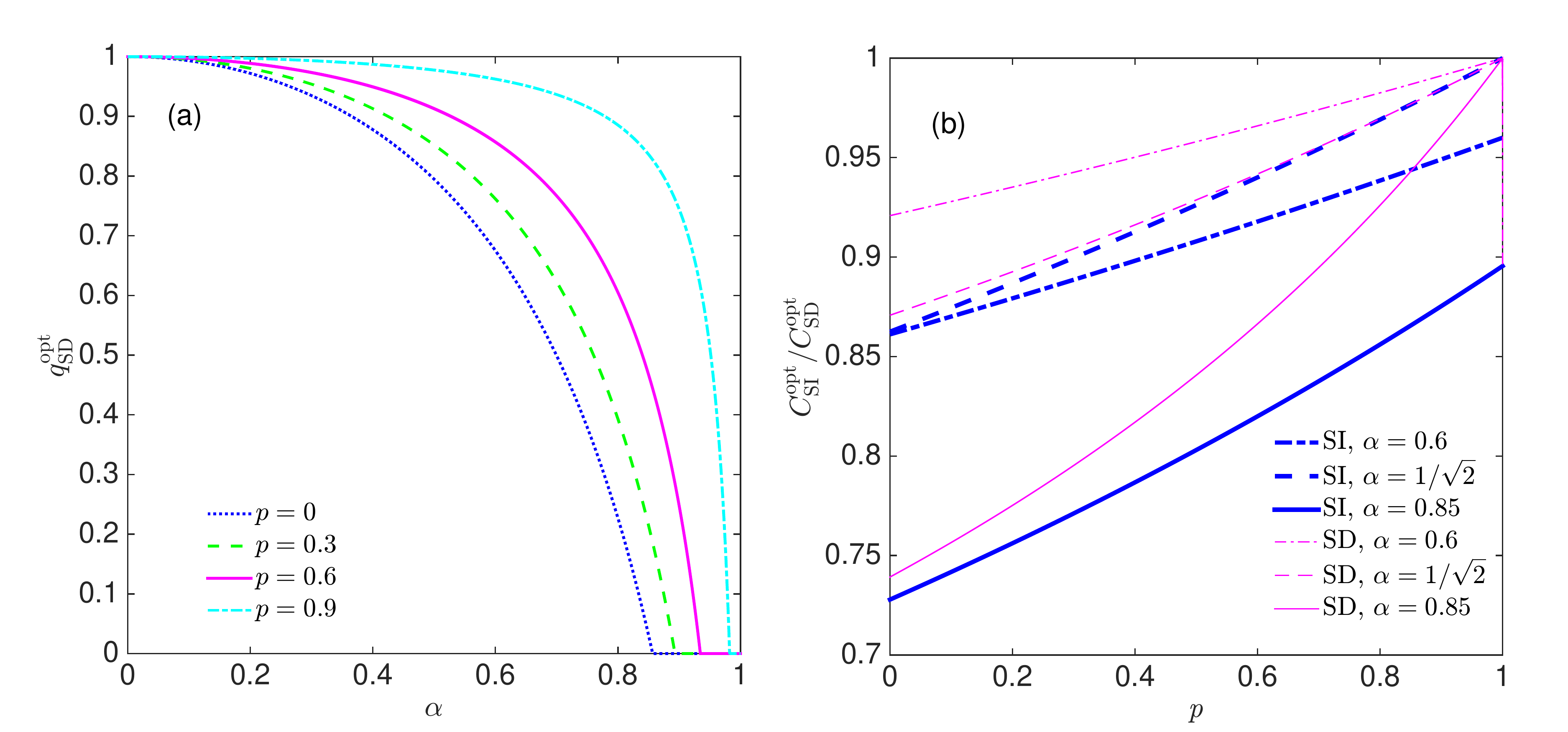}
\caption{(color online) (a). The most optimal reversing measurement strength $q_{\rm SD}^{\rm opt}$ as a function of initial parameter $\alpha$ for different values of $p$ and $r=0.6$. (b) The maximally state-dependent  concurrence $C_{\rm SD}^{\rm opt}$ (thin-mauve lines) as a function of the partial measurement strength $p$ for different initial parameters $\alpha$ and $r=0.6$. For contrast, the maximally state-independent concurrence $C_{\rm SI}^{\rm opt}$ (thick-blue lines) is given as well.}
\label{Fig3}
\end{figure*}

\subsection{State-dependent optimal partial measurement reversal}
\label{sec3.2}
Even though Eq. (\ref{eq3.8}) seems feasible in practice since it is concise and state-independent (i.e., without involving the initial parameters $\alpha$ and $\beta$), it is not the most optimal one in mathematics. Note that one can find the most optimal reversing measurement strength that gives the maximal concurrence by varying $q$ from 0 to 1. Numerical simulation suggests that the most optimal reversing measurement strength is state-dependent, as shown in Fig. \ref{Fig3}(a).

To demonstrate the power of state-dependent optimal partial measurement reversal, we make a comparison between state-dependent  concurrence $C_{\rm SD}^{\rm opt}$ and state-independent concurrence $C_{\rm SI}^{\rm opt}$ in Fig. \ref{Fig3}(b). As can be seen, $C_{\rm SD}^{\rm opt}$ is always larger than $C_{\rm SI}^{\rm opt}$ under the same initial conditions. This improvement is based on the complicated optimization of the reversing measurement strength $q_{\rm SD}^{\rm opt}$, which is usually unpractical in realistic situations. In addition, $q_{\rm SD}^{\rm opt}$ is also dependent on the chosen entanglement measures, e.g., concurrence \cite{wootters1998}, negativity \cite{zyc1998}, logarithmic negativity \cite{plenio2005} or others. Consequently, though the state-dependent optimal partial measurement reversal outperforms the state-independent optimal partial measurement reversal, we prefer to adopt the state-independent method due to its convenience and universality.

Another important phenomenon should be emphasized is that $C_{\rm SD}^{\rm opt}$ always approaches to 1 with $p\rightarrow1$, which is independent of the initial state parameters $\alpha$ and $\beta$. While $C_{\rm SI}^{\rm opt}$ only tends to the initial entanglement $2\alpha\beta$. At the first glance, one might conjecture that entanglement has been created by partial measurement and its reversal, which are local operations. We emphasize that it is just an illusion. From the Eq. (\ref{eq3.7}), we find that the entanglement could not be created by local operations if there is no initial entanglement (i.e., $\alpha=0$ or $\beta=0$). However, initial entanglement can be amplified because of the probabilistic nature of partial measurement \cite{ota2012}. The price we pay for the entanglement amplification is the little success
probability.  Alternatively, this result can be thought as entanglement distillation or entanglement concentration with a single copy of a partially-entangled state \cite{bennett1996,kwiat2001,koashi1999}.

\section{Retrieving the quantum discord}
\label{sec4}
As an important measure of the correlations present in a multipartite quantum state,
entanglement plays an essential role in both fundamental physics and quantum information processing.
In the above analyses, we have focused on retrieving the lost entanglement under Unruh decoherence.
However, other correlation measures such as Bell nonlocality, quantum discord, quantum deficit and measurement-induced disturbance are also widely used to
measure the quantum correlations (please see more details in review \cite{modi2012}). Particularly, the quantum discord usually reveals intricate and subtle behavior since it characterizes all the intrinsic non-classical correlations which may also be contained in separable states. In the following, we show that the technique of partial measurement and its reversal is also valid for recovering the lost quantum discord in non-inertial frames.

\begin{figure}[hbtp] \centering
\includegraphics[width=0.5\textwidth]{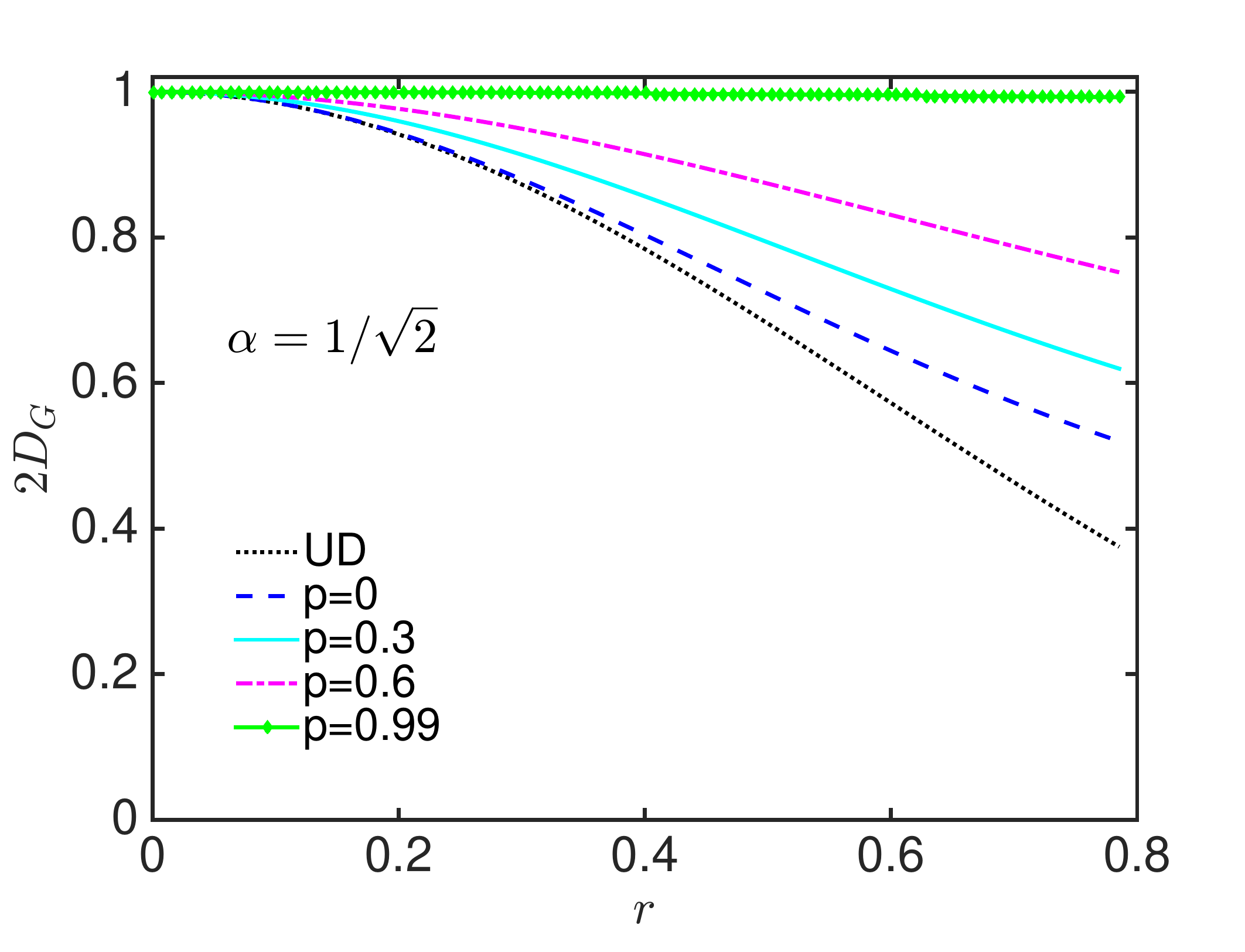}
\caption{(color online) Quantum discord as a function of the acceleration parameter $r$ with $\alpha=1/\sqrt{2}$. The UD line denotes the quantum discord under pure Unruh decoherence without partial measurement. The other lines correspond to $p=0$, $p=0.3$, $p=0.6$ and $p=0.99$, where the state-independent optimal reversing measurement strength $q_{\rm SI}=1-(1-p)\cos^2r$ is used. }
\label{Fig4}
\end{figure}

The quantum discord is usually defined as a difference between total correlations (i.e., quantum mutual information) and the classical correlations. Yet this definition is not convenient for computation since it involves the maximization of the classical correlations \cite{huang2014}. Here, we choose the geometric measure of quantum discord which could be calculated analytically. The state of Eq. (\ref{eq3.5}) can be expressed in the Bloch basis, 
\begin{eqnarray}
\label{eq4.1}
\rho_{A,\rm I}=&&\frac{1}{4}\Big(I_{2}\otimes I_{2}+\sum_{i=1}^{3}x_{i}\sigma_{i}\otimes I_{2}+\sum_{j=1}^{3}y_{j}I_{2}\otimes\sigma_{j}\\
&&+\sum_{i,j=1}^{3}W_{ij}\sigma_{i}\otimes\sigma_{j}\Big),\nonumber
\end{eqnarray}
where $\vec{x}=(0,0,1-2\rho_{44})$ and $\vec{y}=(0,0,2\rho_{11}-1)$ are the Bloch vectors associated to the subsystems A, \rm I, and $W=\rm diag(2\rho_{14},-2\rho_{14},1-2\rho_{22})$ denotes the correlation matrix. $\sigma_{i}$ ($i = 1, 2, 3$) are the Pauli matrices and $I_{2}$ is a identity matrix with two dimension. 
Then, the geometric discord equals \cite{dakic2010}
\begin{equation}
\label{eq4.2}
D_{G}=\frac{1}{4}\left(||\vec{x}\vec{x}^{T}||_{2}+||W||_{2}^{2}-\lambda\right),
\end{equation}
with $\lambda$ being the largest eigenvalue of the matrix $\vec{x}\vec{x}^{T}+WW^{T}$.
Since the maximum value of $D_{G}$ is 1/2 for two-qubit states, so it is natural to consider $2D_{G}$ as a proper measure.

Figure \ref{Fig4} shows how the partial measurement changes the geometric quantum discord. For each partial measurement strength $p$, the state-independent optimal reversing measurement strength $q_{\rm SI}=1-(1-p)\cos^2r$ was chosen.  
The monotonous decrease of $2D_{G}$ as the acceleration increases means that the quantum discord decreases due to the thermal fields generated by the Unruh effect \cite{qiang2015}. However, it is clear that the lost quantum discord can be retrieved by exploiting partial measurement and partial measurement reversal. Particularly, when $p=0.99$, the quantum discord is almost completely recovered even under severe Unruh decoherence, in the expense of very low probability of success.

\section{Conclusions}
\label{sec5}
In summary, we have demonstrated that partial measurement and its reversal can indeed be useful for battling against the Unruh decoherence. The results show that the technique of partial measurement can not only retrieve the lost entanglement shared by an inertial and an accelerated observer but also may amplify it on the basis of adequate optimization of the reversed measurement strength. In view of the probabilistic nature of partial measurement, we attribute this entanglement amplification to the entanglement distillation with a single copy of a partially-entangled state. 

Even though we have investigated the recovery of the lost ferimonic entanglement between an inertial and a noninertial observer, the generalization to the situations of retrieving multipartite entanglement and two uniformly accelerated observers is straightforward. On the other hand,
our method could be extended to recover other types of quantum correlations, such as quantum discord. As a matter of fact, the state-independent strategy is not restricted to the specific quantity of entanglement or quantum discord,  hence it can be exported to a huge variety of physical quantities, for instance, quantum coherence \cite{wang2016a} or quantum Fisher information \cite{yao2014} under Unruh decoherence. The later aspect is of both theoretical and applied significance as the maintenance of quantum Fisher information is a central problem in relativistic quantum metrology \cite{bruschi2014,ahmadi2014} and relativistic clock synchronization \cite{wang2016b,lock2016}.

Note that in this work, we have considered a highly idealized scenario in which the Minkowski annihilation operator is assumed to be one of the right moving Unruh modes ( usually known as the SMA). Under this approximation, the influence of Unruh effect is equivalent to an anti-amplitude damping process, which could be eliminated by the partial measurement and its reversal. In general situations, the SMA may be not valid and some subtle phenomena stem beyond the SMA \cite{bruschi2010,friis2011,montero2011b}. For this case, the Unruh effect can not be expressed as a simple decoherence process. A more artful type of partial measurement needs to be designed. Further work in this respect is underway

\begin{acknowledgements}
This work is supported by the Funds of the National Natural Science
Foundation of China under Grant Nos. 11665004, 11365011, 11605166, 61765007, and supported by Scientic Research Foundation of Jiangxi Provincial Education Department under Grants No. GJJ150996.
\end{acknowledgements}

\appendix
\section{Derivation of the state-independent optimal partial measurement reversal}
To find the state-independent optimal partial measurement reversal, we start with the initial state (\ref{eq3.1}) shared by Alice and Rob. The process contains the following steps: (1) Rob acts a partial measurement on his particle with the measurement strength $p$. (2) Rob moves with constant acceleration. (3) Rob performs out a bit-flip operation in the region \rm I. (4) A second partial measurement is played by Rob with the strength $q$. (5) A final bit-flip is done by Rob. The mathematical formulation is as follows (without normalization):
\begin{eqnarray}
&&\alpha|0\rangle_{A}|0\rangle_{R}+\beta|1\rangle_{A}|1\rangle_{R},\nonumber\\
\xrightarrow{(1)}&&\alpha\sqrt{\overline{p}}|0\rangle_{A}|0\rangle_{R}+\beta|1\rangle_{A}|1\rangle_{R}, \\
\label{eqA1}
\xrightarrow{(2)}&&\alpha\sqrt{\overline{p}}\left(\cos r|0\rangle_{A}|0\rangle_{\rm I}|0\rangle_{\rm II}+\sin r|0\rangle_{A}|1\rangle_{\rm I}|1\rangle_{\rm II}\right)\\
&&+\beta|1\rangle_{A}|1\rangle_{\rm I}|0\rangle_{\rm II},\nonumber\\
\label{eqA2}
\xrightarrow{(3)}&&\alpha\sqrt{\overline{p}}\left(\cos r|0\rangle_{A}|1\rangle_{\rm I}|0\rangle_{\rm II}+\sin r|0\rangle_{A}|0\rangle_{\rm I}|1\rangle_{\rm II}\right)\\
&&+\beta|1\rangle_{A}|0\rangle_{\rm I}|0\rangle_{\rm II},\nonumber\\
\label{eqA3}
\xrightarrow{(4)}&&\alpha\sqrt{\overline{p}}\big(\cos r|0\rangle_{A}|1\rangle_{\rm I}|0\rangle_{\rm II}+\sqrt{\overline{q}}\sin r|0\rangle_{A}|0\rangle_{\rm I}|1\rangle_{\rm II}\big)\nonumber\\
&&+\beta\sqrt{\overline{q}}|1\rangle_{A}|0\rangle_{\rm I}|0\rangle_{\rm II},\\
\label{eqA4}
\xrightarrow{(5)}&&\alpha\sqrt{\overline{p}}\big(\cos r|0\rangle_{A}|0\rangle_{\rm I}|0\rangle_{\rm II}+\sqrt{\overline{q}}\sin r|0\rangle_{A}|1\rangle_{\rm I}|1\rangle_{\rm II}\big)\nonumber\\
&&+\beta\sqrt{\overline{q}}|1\rangle_{A}|1\rangle_{\rm I}|0\rangle_{\rm II}.
\label{eqA5}
\end{eqnarray} 
In order to retrieve the entanglement between Alice and Rob, we expect the optimal $q$ makes the state of 
eq. (\ref{eqA5}) reduces to $\alpha|0\rangle_{A}|0\rangle_{\rm I}+\beta|1\rangle_{A}|1\rangle_{\rm I}$. Remarkably, if we choose 
\begin{equation}
\label{eqA6}
q=1-\overline{p}\cos^{2}r
\end{equation}
then the above eq. (\ref{eqA5}) could be rewritten as 
\begin{eqnarray}
\label{eqA7}
\frac{1}{\sqrt{1+\alpha^2\overline{p}\sin^{2}r}}\Big[\big(&&\alpha|0\rangle_{A}|0\rangle_{\rm I}+\beta|1\rangle_{A}|1\rangle_{\rm I}\big)|0\rangle_{\rm II}\\
&&+\alpha\sqrt{\overline{p}}\sin r|0\rangle_{A}|1\rangle_{\rm I}|1\rangle_{\rm II}\Big].\nonumber
\end{eqnarray}
It is interesting to find that the state of (\ref{eqA3}) becomes closer to $\alpha|0\rangle_{A}|0\rangle_{\rm I}+\beta|1\rangle_{A}|1\rangle_{\rm I}$ with the increase of measurement strength $p$. Particularly, when $p=1$, the initial entangled state between Alice and Rob is recovered. In this case, the optimal reversing measurement strength of eq. (\ref{eqA6}) is irrelevant to the initial parameters $\alpha$ and $\beta$, so we call it state-independent partial measurement reversal.



\end{document}